\documentclass[aps,prl,twocolumn,superscriptaddress]{revtex4}
\usepackage{graphicx,amsmath,amssymb,bm}

\newcommand{\kf}{k_{\rm F}}
\newcommand{\ef}{\varepsilon_{\rm F}}
\newcommand{\as}{a_{\rm s}}
\newcommand{\re}{r_{\rm e}}
\newcommand{\be}{\begin{equation}}
\newcommand{\ee}{\end{equation}}
\newcommand{\fm}{\, \text{fm}}
\newcommand{\fmi}{\, \text{fm}^{-1}}

\newcommand{\vlk}{V_{{\rm low}\,k}}
\newcommand{\la}{\Lambda}
\begin{document}

\title{Resonant Fermi gases with a large effective range}
\author{A. Schwenk}
\email[E-mail:~]{schwenk@indiana.edu}
\affiliation{Nuclear Theory Center, Indiana University,
Bloomington, IN 47408}
\author{C.J.  Pethick}
\email[E-mail:~]{pethick@nordita.dk}
\affiliation{NORDITA, Blegdamsvej 17, DK-2100 Copenhagen \O, Denmark}


\begin{abstract}

We calculate the equation of state of a Fermi gas with resonant
interactions when the effective range is appreciable.  Using an
effective field theory for large scattering length and large effective
range, we show how calculations in this regime become
tractable.  Our results are model-independent, and as an application,
we study the neutron matter equation of state at low densities of 
astrophysical interest $0.002 \fm^{-3} < \rho < 0.02 \fm^{-3}$, for 
which the interparticle separation is comparable to the effective range.
We compare our simple results with those of conventional
many-body calculations.

\end{abstract}

\pacs{03.75.Ss, 05.30.Fk, 21.65.+f}
\keywords{Resonant Fermi gases, low-density neutron matter, many-body theory}

\maketitle

The properties of dilute Fermi gases when all two-body scattering
lengths $\as$ are large compared to the interparticle separation
$r_{\rm s}$ and the range of the interaction $R$ is small compared 
with $r_{\rm s}$ are universal, in the sense that properties do not
depend on details of the interparticle interaction~\cite{BBH}. 
This is because the only
dimensionful scale is the Fermi momentum $\kf$, and the corresponding energy
scale is the Fermi energy $\ef = \kf^2/(2m)$, $m$ being the fermion mass.
Consequently all macroscopic observables
are given by appropriate powers of $\kf$ or $\ef$ multiplied by
universal factors. For instance, for two spin states with equal
populations the 
energy per particle $E/N$ of cold gases of, e.g., $^6$Li, $^{40}$K atoms 
or neutrons under these conditions is given by
\be
\frac{E}{N} = \xi \biggl( \frac{E}{N} \biggr)_{\rm free} = \xi \,
\frac{3 \kf^2}{10 m} \,,
\label{unieos}
\ee
where the universal factor $\xi$ is a number.

Resonant, dilute Fermi gases were realized for the first time
by O'Hara {\it et al.}~in 2002~\cite{Thomas1}.
The factor $\xi$ has been determined experimentally by
a number of methods~\cite{Thomas1,Thomas2,Salomon,Grimm}
in atomic gases where the interaction may be tuned by
controlling the magnetic field. The
extraction of the equation of state leads to
$\xi = 0.51 \pm 0.04$~\cite{Thomas2}, $\xi \approx 0.7$~\cite{Salomon}
and $\xi = 0.27\,^{+0.12}_{-0.09}$~\cite{Grimm}, for temperatures in units
of the Fermi temperature $T/T_{\rm F}
\approx 0.05$ (except for $T/T_{\rm F} \approx 0.6$ in~\cite{Salomon}).
To date, the most reliable theoretical results are from $T=0$,
fixed-node Green's function Monte Carlo simulations,
$\xi = 0.44 \pm 0.01$~\cite{Carlson} and
$\xi = 0.42 \pm 0.01$~\cite{Astra}.

The purpose of this letter is to consider the neutron gas as an
example of a Fermi system with a large scattering length.
For neutrons, the scattering length is unnaturally large and the
currently accepted value is $a_{\rm nn} = - 18.5 \pm 0.3 \fm$
(for a recent review on the experimental situation see~\cite{Phillips}).
This is to be compared with the range of nuclear interactions, which is
given by the mass of the lightest exchange particle, the pion, and thus
$R \sim 1/m_\pi \approx 1.4 \fm$. The effective
range $\re$ is expected to be approximately charge independent, and therefore
we take the neutron-neutron effective range to be $r_{\rm nn} = 2.7
\fm$, the same as the neutron-proton one~\cite{effrange}.
Consequently, the neutron effective range is also
significant, with $\re m_\pi \approx 2$.

For neutrons, the validity of the universal equation of state applies
to densities with $\kf \re \ll 1$. This restricts neutron
densities to
$\rho = \kf^3/(3\pi^2) <
10^{-4} \fm^{-3}$, which is just below neutron drip density
$\rho_{\rm nd} = 2.3 \cdot 10^{-4} \fm^{-3} \sim 10^{-3}
\, \rho_0$ at which neutrons become
unbound. Here $\rho_0 = 0.16 \fm^{-3}$ is the saturation density
of symmetric nuclear matter.
Therefore, it is important to generalize the equation of state
to resonant Fermi gases with an appreciable effective range. In the
regime $\kf \re \sim 1$, the energy per particle can be expressed as
in Eq.~(\ref{unieos}) but with a system-specific factor $\xi(\kf \re)$,
\be
\frac{E}{N} = \xi(\kf \re) \, \frac{3 \kf^2}{10 m} \,.
\label{larger}
\ee
This is the case for neutron matter at subnuclear densities $\rho <
\rho_0/10$.
For the extension Eq.~(\ref{larger}) to be valid, the
contributions to the energy from higher partial waves and many-body
forces must be negligible. This ensures that neutron densities with
$\kf \re \sim 1$ do not probe further details of nuclear forces.
In addition, the effective range expansion truncated at $\re$,
$k \cot\delta(k) = -1/\as + \re k^2/2$, must describe the
phase shifts $\delta(k)$ for S-wave scattering
up to relative momenta $k \sim \kf$. The extension of the
universal energy to $\kf \re \sim 1$ is also valid for trapped
Fermi gases where both the scattering length and the effective
range are tuned to large values. In this letter, we calculate
$\xi(\kf \re)$ for resonant Fermi gases with $\kf \re \gtrsim 1$
and explain how this regime becomes theoretically tractable.

\begin{figure}
\begin{center}
\includegraphics[scale=0.4,clip=]{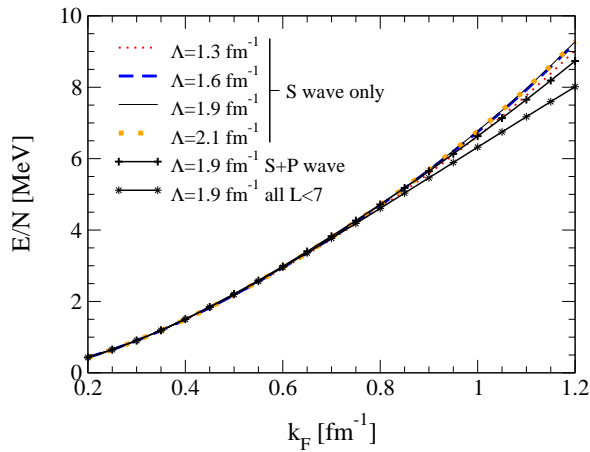}
\end{center}
\vspace*{-4mm}
\caption{\label{hf} Hartree-Fock results for the energy per particle
of neutron matter calculated from $\vlk$ for different cutoffs.}
\end{figure}

For neutron matter, the non-S-wave contributions to the energy
per particle can be estimated from the Hartree-Fock results
shown in Fig.~\ref{hf}. The Hartree-Fock calculations
are based on the model-independent two-nucleon interaction $\vlk$ for
different values of the momentum cutoff $\Lambda$. The evolution
of $\vlk$ with the cutoff follows a renormalization group (RG)
equation that guarantees that nucleon-nucleon phase shifts
are cutoff independent. We have previously found that the evolution
of all large-cutoff nuclear interactions to $\la \lesssim 2.1 \fmi$
leads to the same low-momentum interaction $\vlk$~\cite{Vlowk}.
Varying the cutoff gives an approximate measure of higher-order
contributions beyond Hartree-Fock and the effects of omitted
many-body forces, since both are needed to obtain cutoff-independent
results. We find in Fig.~\ref{hf} that the low-density equation of
state is cutoff-independent for $\kf \lesssim 1.0 \fmi$, with
negligible $l \geqslant 1$ contributions for $\kf \lesssim 0.8 \fmi$.
In addition, the effective range expansion
describes the $^1$S$_0$ phase shifts well up to relative
momenta $k < 1.0 \fmi$, with a deviation 
of $-0.4$, $0.2$ and $2.1$ degrees from the empirical phase
shifts $57.7$, $48.6$ and $39.5$ degrees for $k = 0.4$, $0.6$ and
$0.8 \fmi$ respectively. Consequently, the equation of state of
neutron matter is given by Eq.~(\ref{larger}) for $\kf \re \lesssim 2$.

We calculate the equation of state Eq.~(\ref{larger}) using an
effective field theory (EFT) for large scattering length and large
effective range. EFT offers a systematic approach to interactions
at low energies, and therefore our results are model-independent.
Under the conditions of interest, both the
scattering length and effective range are low-momentum scales
$1/\as \sim 1/\re \sim Q$. The corresponding EFT is realized by
introducing a di-fermion field $d$~\cite{KBS} (for a review see
Ref.~\cite{Daniel}) with the lowest-order Lagrangian density given
by (in units $\hbar=m=1$)
\begin{eqnarray}
{\mathcal L} &=& \psi^\dagger \biggl( i \partial_0 + \frac{\nabla^2}{2}
\biggr) \, \psi - d^\dagger \biggl( i \partial_0 + \frac{\nabla^2}{4}
- \Delta \biggr) \, d \nonumber \\[1mm]
&-& g \, \bigl( d^\dagger \psi\psi + d \, \psi^\dagger \psi^\dagger \bigr) \,.
\label{dEFT}
\end{eqnarray}
Here $\psi$ denotes the fermion field, and $\Delta$ and $g$ are
low-energy constants which describe the propagation of the di-fermion
field and its coupling to two fermions respectively. We also note
that $i \partial_0 + \nabla^2/4$ is the operator for the
two-body energy corrected for the center-of-mass motion. The EFT
Lagrangian Eq.~(\ref{dEFT}) was introduced for nucleon-nucleon scattering in
Ref.~\cite{KBS} following work due to Weinberg~\cite{Weinberg}.
In the di-fermion EFT, the
$T$ matrix for two-body scattering depends only on the
energy $E$ of the interacting particles in the center-of-mass system,
\be
T(E) = \frac{g^2}{\Delta - E - g^2 \, I_0(E+i\epsilon)} \,,
\ee
with $I_0(E) = \int_0^\la d^3p/(2\pi)^3 \, (E-p^2)^{-1}$.
This corresponds to summing diagrams where the di-fermion
propagator (represented as a double line) is dressed by
fermion loops,
\begin{widetext}\vspace*{-6mm}\be
\includegraphics[scale=0.85,clip=]{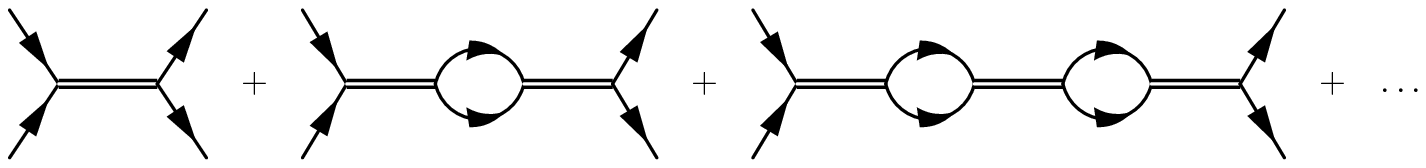}
\label{tdiags}
\ee\end{widetext}
$I_0(E)$ denotes the loop integral regulated by a momentum
cutoff $\la$ and the on-shell $T$
matrix is given by $4\pi \, T^{-1}(E=k^2) = 1/\as - \re \, k^2/2 + i k$,
which can equivalently be obtained from an
energy-dependent potential,
\be
V(E) = \frac{g^2}{\Delta - E} \,,
\ee
which follows after integrating out the di-fermion field.
Matching the lowest-order low-energy constants $\Delta$
and $g$ to the effective range expansion yields
\be
\frac{\Delta}{g^2} = \frac{1}{4 \pi \as} - \frac{\la}{2 \pi^2}
\quad \text{and} \quad
\frac{1}{g^2} = \frac{\re}{8\pi} - \frac{1}{2 \pi^2 \, \la} \,.
\ee
Consequently, for systems with large scattering length and
large effective range the potential $V(E)$ scales at low
energies $E \sim Q^2$ as $V(E) \sim (Q + \la + Q^2/\la)^{-1}$.
The cutoff generates higher-order terms
$\sim k^4/\la^3$ in $T^{-1}(E=k^2)$, which are suppressed
for $k/\la \ll 1$. Therefore, we consider the lowest-order
EFT with large cutoffs. This is not required for $\vlk$,
since the RG generates all higher-order contact interactions
necessary to maintain cutoff-independent two-body observables.
If one rescales the di-fermion field by $d \to g \, d$, the
interaction terms become coupling-independent and the kinetic
term acquires a factor $1/g^2$. For large cutoffs $1/g^2 \sim \re$,
and therefore the di-fermion field is an auxiliary field for
positive effective range, even if there is a two-body bound
state with $\as > 0$. Integrating out the $d$ field gives the
standard contact interaction EFT. For negative effective range
the di-fermion kinetic term has the normal sign and the
Lagrangian Eq.~(\ref{dEFT}) is the atom-molecule model.
The great advantage of the present formulation is that it
enables one to study systems with either positive or negative $\re$.

Next, we study the scaling of diagrammatic contributions to
the energy. Particle-particle loops are restricted only by
the cutoff, and therefore these scale as $\sim \la$ for large
cutoffs. By contrast, particle-hole and hole-hole intermediate
states scale with the low-momentum scale~$\sim Q$. Particle-hole
loops, for instance, scale with the density of states $\sim \kf$.
Since the lowest-order interaction scales as $V(E) \sim
(Q + \la + Q^2/\la)^{-1}$, cutoff independence for large
cutoffs minimally requires summing particle-particle ladders with
Pauli blocking, such that the $\la^{-1}$ scaling of the potential
cancels with each loop $\sim \la$~\cite{Vlowkcomment}. Neglecting
terms which vanish for large cutoffs, we find for $T_{\text{med}}$, the
on-shell in-medium $T$ matrix
\begin{widetext}\vspace*{-6mm}\be
T_{\text{med}}(E=k^2;P) = 4\pi \biggl[ \frac{1}{\as} - \frac{\re \, k^2}{2}
- \frac{\kf + \frac{P}{2}}{\pi}
+ \frac{k}{\pi} \, 
\log\biggl(\frac{\kf+\frac{P}{2}+k}{\kf+\frac{P}{2}-k}\biggr)
+ \frac{k^2+\frac{P^2}{4}-\kf^2}{\pi \, P} \,
\log\biggl(\frac{(\kf+\frac{P}{2})^2-k^2}{\kf^2
-\frac{P^2}{4}-k^2}\biggr) \biggr]^{-1} \,,
\label{tmed}
\ee\end{widetext}
where P denotes the particle-pair momentum and we have neglected
the imaginary part, as the latter vanishes for $k < \sqrt{\kf^2-P^2/4}$
and therefore does not contribute to the energy. The resulting
in-medium $T$ matrix can be viewed as an effective
interaction and scales as $T_{\text{med}}(E=k^2;P) \sim Q^{-1}$
independent of the cutoff. In addition to the scattering length
and effective range terms, the inverse effective interaction acquires
a density-dependent contribution $-(\kf+P/2)/\pi$ and two logarithmic
terms. The first logarithmic term includes the BCS singularity
for $P=0$ and $k=\kf$. The resulting contribution to the energy
is integrable and generally small. For  neutron matter, realistic
calculations of the superfluid pairing gap including polarization
effects give $\Delta/\ef \sim 0.1$ for densities with $\kf \re
\sim 1$~\cite{RGnm}. We therefore expect corrections to the energy
per particle $\xi_{\text{sf}}(\kf \re) \approx -5/8 \, (\Delta/\ef)^2
\approx - 0.01$ due to pairing in this regime. The second logarithmic
term in Eq.~(\ref{tmed}) is regular, and for $P=0$ it simplifies to $-\kf/\pi$.

For large effective range, interactions are weaker at higher momenta
and the effect on the in-medium $T$ matrix can be
understood qualitatively by inserting average values for
the momenta $k^2 = 3/10 \, \kf^2$ and $P^2 = 6/5 \, \kf^2$ into
Eq.~(\ref{tmed}). With this qualitative estimate, the
average effective interaction for resonant Fermi gases 
is given by
\be
\overline{T_{\text{med}}} = - \, \cfrac{4\pi}{U_0 \, \kf + \cfrac{3 \re 
\kf^2}{20}} =
- \, \frac{4\pi}{U_0 \, \kf} \, \frac{1}{1 + 0.27 \, \kf \re} \,,
\label{avint}
\ee
with $U_0 = 0.56$.
In the universal regime with $\kf \re \ll 1$,
the average effective interaction is given by $\overline{T_{\text{med}}}
= - 4\pi/(U_0 \kf)$. The remaining contributions to the energy are of
particle-hole or hole-hole nature, and scale as $Q \sim \kf$.
In this EFT it is thus necessary to sum all diagrams, as
the effects of each additional vertex $\sim \kf^{-1}$ combined with
the additional loop $\sim \kf$ are of order one.

\begin{figure}
\begin{center}
\includegraphics[scale=0.5,clip=]{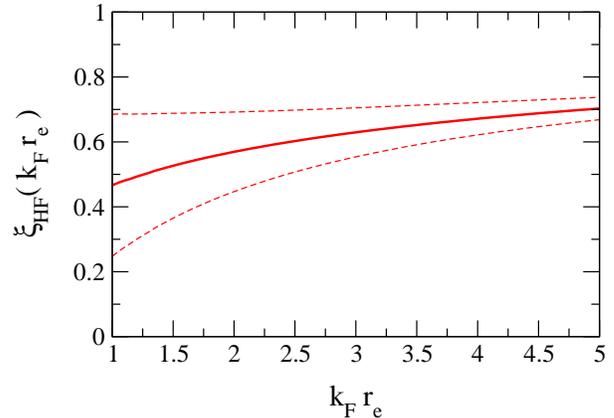}
\end{center}
\vspace*{-6mm}
\caption{\label{xi} Result for $\xi_{\text{HF}}(\kf \re)$ (solid
line) which describes the equation of state of resonant Fermi gases
with large effective range $\kf \re \gtrsim 1$. The dashed lines provide
an error estimate for $\xi_{\text{HF}}$ as discussed in the text.}
\end{figure}

For Fermi gases with large effective range, the loop scaling is unchanged
but additional $T_{\text{med}}$ vertices are now suppressed by
$(1 + 0.27 \, \kf \re)^{-1}$.
This implies that corrections to the Hartree-Fock equation of state
with $T_{\text{med}}$ as effective interaction come with powers of
$(1 + 0.27 \, \kf \re)^{-1}$ relative to the universal case.
In perturbation theory, particle-particle-irreducible contributions
to the energy enter in third order. With $\overline{T_{\text{med}}}
= 4 \pi \, \overline{a_{\text{med}}}$, we estimate the leading non-ladder
corrections to the $\xi(\kf \re)$ obtained from $T_{\text{med}}$
at the Hartree-Fock level as $\Delta \xi = 0.077 \, 
(U_0 \, (1 + 0.27 \, \kf \re))^{-3}$~\cite{Baker}. This only provides a
qualitative estimate, since we evaluated the in-medium $T$
matrix for average particle configurations.

To a good approximation, the equation of state of resonant Fermi gases
with $\kf \re \gtrsim 1$ can therefore be calculated from $T_{\text{med}}$
at the Hartree-Fock level,
\begin{eqnarray}
\xi_{\text{HF}}(\kf \re)
&=& 1 + \frac{5}{2\pi^2 \, \kf^5} \int\limits_0^{2\kf} P^2 \, dP
\int\limits_0^{\sqrt{\kf^2-\frac{P^2}{4}}} k^2 \, dk \nonumber \\[1mm]
&\times& T_{\text{med}}(k^2;P)
\min\biggl[1 , \frac{\kf^2-k^2-\frac{P^2}{4}}{k \, P}\biggr] , \quad
\label{xihf}
\end{eqnarray}
with $1/\as=0$ in Eq.~(\ref{tmed}).
Based on the arguments given above, we expect corrections to
$\xi$ beyond Hartree-Fock of order $\Delta \xi = 0.22 - 0.03$
and $\xi_{\text{sf}} \approx - 0.01$ due to pairing for densities
with $\kf \re = 1 - 5$. Our results for $\xi_{\text{HF}}(\kf
\re)$ are shown in Fig.~\ref{xi} with the theoretical error
estimate. It is intriguing that we find $\xi_{\text{HF}}(1)
\approx 0.47$, which is close to the universal factor
$\xi=0.44 \pm 0.01$~\cite{Carlson}. This implies a cancellation
of higher-order contributions for $\kf \re \ll 1$.

For $\kf \re \gg 1$, the effective interaction is of the order
$\overline{T_{\text{med}}} \sim - (\re \kf^2)^{-1}$ and
interaction effects disappear for infinite effective range.
In the latter regime, $\xi \approx 1 + C/(\kf \re)$, with
a coefficient $C < 0$ of order one. Consequently, $\xi \to 1$
only for very large effective ranges. We find $\xi_{\text{HF}}(10)
= 0.79$ and $\xi_{\text{HF}}(100) = 0.96$. If the effective
range is negative, the average interaction in the medium is
greater than for $\re = 0$, and higher-order terms are only
reduced for very large $|\kf \re| \gg 1$.

\begin{figure}
\begin{center}
\includegraphics[scale=0.5,clip=]{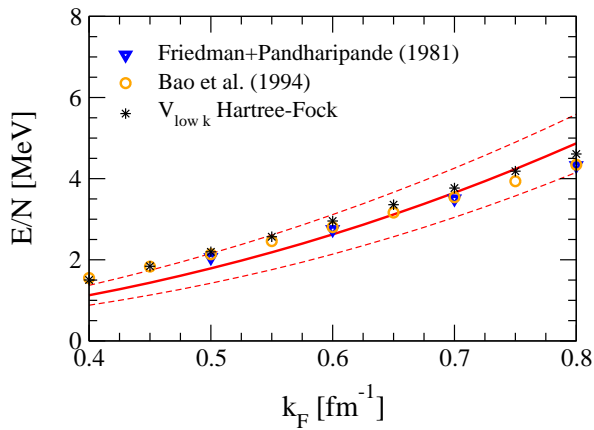}
\end{center}
\vspace*{-6mm}
\caption{\label{comp} Comparison of the energy per particle for
neutron matter calculated from Eq.~(\ref{xihf}) for $\as = -18.5 \fm$
and $\re = 2.7 \fm$ (solid line) to Fermi hypernetted-chain
(triangles)~\cite{FP} and Brueckner results (circles)~\cite{Bao}. 
We also show the $\vlk$ Hartree-Fock
results of Fig.~\ref{hf}. The dashed lines provide an error estimate
for $\xi_{\text{HF}}$.}
\end{figure}

Finally, we take into account a large, but finite scattering length,
and present in Fig.~\ref{comp} results for the equation of state of
neutron matter in the regime $\kf \re \sim 1$.
This is straightforward by including the $1/\as$ term in Eq.~(\ref{tmed}).
Our results are model-independent and constrain neutron matter
at subnuclear densities $0.002 \fm^{-3} < \rho < 0.02 \fm^{-3}$,
which is of relevance to the physics of neutron stars and
supernovae. We also compare our neutron matter results to
many-body calculations using Fermi hypernetted-chain
techniques~\cite{FP} or Brueckner theory~\cite{Bao}.
As can be seen from Fig.~\ref{comp} all microscopic results
lie within our theoretical error estimates.

In summary, we have calculated the equation of state of resonant
Fermi gases with large effective range $\kf \re \gtrsim 1$ in the
di-fermion EFT, with particular attention to neutron matter at
subnuclear densities. In the lowest-order EFT, cutoff independence
minimally requires the resummation of particle-particle ladders~[13],
which leads to an effective interaction that becomes weak for all
$\kf \re \gtrsim 1$. The $\kf \re$ dependence was used to show
how resonant Fermi gases with large effective range, such as
low-density neutron matter, are theoretically tractable and to
estimate the error of our results. Neutron matter equations
of state obtained using conventional many-body approaches are
consistent with our model-independent di-fermion EFT results
within these errors. For $\kf \re \sim 1$, the energy is found
to be close to the universal regime.

\vspace*{-4mm}

We thank Georg Bruun, Dick Furnstahl, 
Chuck Horowitz and Daniel Phillips for useful
discussions. The work of AS is supported by the US DOE
under Grant No. DE--FG02--87ER40365 and the NSF
under Grant No. PHY--0244822.


\vspace*{-2mm}

\begin{thebibliography}{99}
\bibitem{BBH} G.F. Bertsch posed the properties of Fermi gases in 
this limit as a many-body-theory challenge problem at MBX (2001). The
first studies were reported by G.A. Baker, Int. J. Mod. Phys. \textbf{B15},
1314 (2001); H. Heiselberg, Phys. Rev. \textbf{A63}, 043606 (2001).
\bibitem{Thomas1} K.M. O'Hara, S.L. Hemmer, M.E. Gehm, S.R. Granade
and J.E. Thomas, Science \textbf{298}, 2179 (2002).
\bibitem{Thomas2} M.E. Gehm, S.L. Hemmer, S.R. Granade, K.M. O'Hara
and J.E. Thomas, Phys. Rev. \textbf{A68}, 011401(R) (2003); J. Kinast,
A. Turlapov, J.E. Thomas, Q. Chen, J. Stajic and K. Levin, Science
\textbf{307}, 1296 (2005).
\bibitem{Salomon} T. Bourdel, J. Cubizolles, L. Khaykovich, K.M.F.
Magalhaes, S. Kokkelmans, G.V. Shlyapnikov and C. Salomon, Phys.
Rev. Lett. \textbf{91}, 020402 (2003).
\bibitem{Grimm} M. Bartenstein, A. Altmeyer, S. Riedl, S. Jochim,
C. Chin, J.H. Denschlag and R. Grimm, Phys. Rev. Lett. \textbf{92},
120401 (2004) and cond-mat/0412712.
\bibitem{Carlson} J. Carlson, S.-Y. Chang, V.R. Pandharipande and
K.E. Schmidt, Phys. Rev. Lett. \textbf{91}, 050401 (2003);
S.Y. Chang, J. Carlson, V.R. Pandharipande and K.E. Schmidt,
Phys. Rev. \textbf{A70}, 043602 (2004).
\bibitem{Astra} G.E. Astrakharchik, J. Boronat, J. Casulleras and
S. Giorgini, Phys. Rev. Lett. \textbf{93}, 200404 (2004).
\bibitem{Phillips} A. Gardestig and D.R. Phillips, nucl-th/0501049.
\bibitem{effrange} The
neutron-proton effective range parameters are well-constrained
from the Nijmegen 1993 partial-wave analysis (PWA93): 
$a_{\rm np} = - 23.768 \pm 0.006 \fm$ 
and $r_{\rm np} = 2.68 \pm 0.01 \fm$ (M.C.M. Rentmeester, 
private communication).
\bibitem{Vlowk} S.K. Bogner, T.T.S. Kuo and A. Schwenk, Phys. Rept.
\textbf{386}, 1 (2003).
\bibitem{KBS} D.B. Kaplan, Nucl. Phys. \textbf{B494}, 471 (1997);
S.R. Beane and M.J. Savage, Nucl. Phys. \textbf{A694}, 511 (2001).
\bibitem{Daniel} D.R. Phillips, Czech. J. Phys. \textbf{52}, B49 (2002).
\bibitem{Weinberg} S. Weinberg, Phys. Rev. \textbf{130}, 776 (1963).
\bibitem{Vlowkcomment} This
resummation is not necessary if one evolves the large-cutoff theory to
smaller cutoffs using the RG, see 
S.K. Bogner, A. Schwenk, R.J. Furnstahl and
A. Nogga, Nucl. Phys. \textbf{A} in press, nucl-th/0504043.
\bibitem{RGnm} A. Schwenk, B. Friman and G.E. Brown, Nucl. Phys.
\textbf{A713}, 191 (2003); see also Fig.~2 in
A. Schwenk, nucl-th/0411070.
\bibitem{Baker} The third-order non-ladder contributions to the
energy have been evaluated by Baker, see Eq.~(4.84) in G.A.
Baker, Rev. Mod. Phys. \textbf{43}, 479 (1971). Note that for
$\re=0$, $\overline{T_{\text{med}}}$ leads to $\xi_{\rm HF} = 0.36$ 
and $\Delta \xi = 0.45$.
\bibitem{FP} B. Friedman and V.R. Pandharipande, Nucl. Phys.
\textbf{A361}, 502 (1981).
\bibitem{Bao} G. Bao, L. Engvik, M. Hjorth-Jensen, E. Osnes and
E. {\O}stgaard, Nucl. Phys. \textbf{A575}, 707 (1994).
\end{thebibliography}
\end{document}